# A Metropolis Approach for Mesh Router Nodes placement in Rural Wireless Mesh Networks

Jean Louis Ebongue Kedieng Fendji [1*], Christopher Thron [2], Jean Michel Nlong [3]
[1] University of Bremen, Bremen, Germany.
[2] Texas A&M University Central Texas, Killeen, USA.
[3] University of Ngaoundéré, Ngaoundéré, Cameroon.

* Corresponding author. Tel.: +4915213549359; email: fendjike@uni-bremen.de

**Abstract:** Wireless mesh networks appear as an appealing solution to reduce the digital divide between rural and urban regions. However the placement of router nodes is still a critical issue when planning this type of network, especially in rural regions where we usually observe low density and sparse population. In this paper, we firstly provide a network model tied to rural regions by considering the area to cover as decomposed into a set of elementary areas which can be required or optional in terms of coverage and where a node can be placed or not. Afterwards, we try to determine an optimal number and positions of mesh router nodes while maximizing the coverage of areas of interest, minimizing the coverage of optional areas and ensuring connectivity of all mesh router nodes. For that we propose a particularized algorithm based on Metropolis approach to ensure an optimal coverage and connectivity with an optimal number of routers. The proposed algorithm is evaluated on different region instances. We obtained a required coverage between 94% and 97% and a coverage percentage of optional areas less than 16% with an optimal number of routers $nr_{max-2}=1.3*nr_{min}$, ($nr_{min}$ being the minimum number of router which is the ratio between the total area requiring coverage and the area which can be covered by a router).

## 1. Introduction

Recent advances in wireless networks foster the deployment of wireless mesh networks (WMNs) [1]. These networks are mainly composed of mesh router nodes connected in a mesh topology. Based on Wi-Fi technologies, WMNs show themselves to be an appealing solution to bridge the digital divide observed between rural and urban regions. Especially in developing countries where rural activities like breeding, fishing and agriculture still remain the pillar of the economy, WMNs can play a crucial role in the national development. The success of this kind of network is due to the low cost of the Wi-Fi technology and the continuous capacity improvement of this technology in terms of throughput and transmission range. WMN in rural region is usually composed of one gateway which connects the network to Internet, and a set of mesh routers (MRs). Similar to normal routers, MRs incorporate some functionality to support mesh networking.

The performance in terms of coverage and connectivity of a WMN relies on many factors among which the number and the placement of MRs. This number has a direct incidence on the cost of the architecture, which is the main concern especially in rural regions in developing world. Therefore, finding the optimal number and placement of routers is a crucial concern.

By its nature, the problem of mesh node placement requires a multi objective approach; since it is a combinatorial optimization problem which is hard to solve in polynomial time, especially when considering a large space to cover [2], [3]. Most of the time, these objectives seem to be contradictory like: minimizing the number of MR while keeping or extending the coverage.

In this paper, we address the problem under a constrained network model tied to rural regions where we usually observe low density and sparse population. We first decompose the area to cover into elementary areas which can be required (school, hospital…) or optional (farm, road…) in terms of coverage and where a node can be placed or not. The objectives here are:

1) To minimize the number of MRs by avoiding covering optional areas as much as possible;
2) To maximize the coverage of required areas;
3) To ensure the connectivity of the network.

We firstly define the network model and provide a formulation of the placement problem in rural region. Afterwards, we propose an effective heuristic to obtain a close to optimal coverage of required areas using an optimal number of routers. The algorithm is based on metropolis approach. Finally, we evaluate the proposed algorithm on different regions instances using Scilab 5.4.0.

In this paper, we extend the work done in [4] by clearly defining the approach used to solve the problem and by ensuring the connectivity constraint. We also evaluate the particularized algorithm on different region instances and we try to provide a generalization of results.

The rest of the paper is organized as follows: In Section 2, we briefly present previous work in this topic. In Section 3, we give the network model and a formulation for the placement problem. The metropolis approach for this problem is described in Section 4. In Section 5, we present the experimental setup to evaluate our approach and discuss the results. We finally conclude the paper in Section 6 and we present limitations and future work in Section 7.

## 2. Related Work

The nodes placement problem is a crucial issue not only in a wireless mesh network design [2]-[16] but also in wireless sensor networks [17], [18]. Usually, this problem is assimilated to the problem of facilities and locations where we have a set of facilities (mesh routers in this case) and a set of locations (the region or the universe) and we have to assign or to build each facility to a location while satisfying certain constraints. There exist different formulations of this mesh router nodes placement problem depending on:

- The type of node involved in the placement problem;
- The characterization of the universe;
- The constraints and Multi-objectiveness in optimization.

Depending on the type of node, the problem of placement usually consists of determining either the position of the gateway [13]-[15] or the position of mesh router nodes [9], [10] in partial design. But when designing the network from scratch, this problem consists of finding an optimal location for both gateway and mesh router nodes [6]-[8], [11], [12].

The universe where to deploy facilities can be considered as: continuous that means we have a continuum region where the facilities may be placed; as discrete [8]-[10] that means the universe is composed of predefined positions; or as Network [7] that means the universe is assimilated to an undirected weighted graph.

Most of the problem formulations are multi-objectives or multi-criteria [6]-[11] and these objectives are optimized either hierarchically, in this case a priority is assigned to each objective then they are optimized following this order, or simultaneously, that means the objectives are combined to find a compromise.

To solve this problem, different approaches have been proposed among which: Graph-theoretic-based approaches [7], [8]; Meta-heuristic based approaches [3], [6], [9]-[12] and linear programming based approaches [2].

In [2] optimization models are proposed for placement of MR with the objectives to minimize the network installation cost and at the same time providing full coverage to wireless mesh clients. The approach is based on mixed integer linear programming and allows selecting the number and positions of mesh routers and access points.

In [8], the authors study efficient mesh router nodes placement in WMN. Their MR placement problem is the determination of a minimum set of positions among the candidate positions in such a way that MRs situated in these positions cover the given region while satisfying the traffic constraint.

The more close approach to our work is the one in [3] which proposes a simulated annealing approach to solve the mesh nodes placement problem. It considers the version of the mesh node placement problem where: given a two-dimensional area where to distribute a number of MR nodes and a number of mesh client nodes of fixed positions (of an arbitrary distribution) it has to find a location assignment for the MRs that maximizes the network connectivity and client coverage.

Although there exist several works on mesh router nodes in WMN, there is still a need to consider rural regions where we observe sparse population, where we do not need to cover a whole region, and where we cannot place a router in some areas. So we need a suitable network model tied to rural regions and clear problem statement.

## 3. Network Model and Problem Statement

### 3.1. Network Model

In urban regions, the population is so dense. Therefore, the more the coverage is great, the more the number of clients increases and the network providers can increase their profit. In contrary to urban regions, in rural regions there is no need to cover a whole region. Because the more the region to cover is great, the more we need router nodes and the overall cost is more expensive. A good approach is to focus only on important parts of a given region. Usually, a given region is composed of:

- Areas of interest (AI) which are usually sparse and where the signal must be spread, like a market, a school, a hospital… When an area is not of interest, the coverage of this region is considered as optional. These regions with optional coverage could be great farms for example.
- Prohibited areas (PA) where a node cannot be placed for example a lake, a river or a road... However, a node can be placed in an optional area because an optional area is not necessary a prohibited one.

In rural regions, the network usually contains only one gateway (IGW) generally fixed and connected to Internet by Satellite. We assume that all routers are equipped with omnidirectional antenna and they have all the same coverage. Hence we represent a router as a circle.

To be realistic, the area to cover is modelled as a two-dimensional irregular form in a two-dimension coordinate plane. We consider the smallest rectangle that can contain the irregular form. Therefore, we assume that this rectangle is decomposed in small square forms called elementary area (EA) in other to obtain a grid. Hence, we obtain a set of elementary areas of interest (EAI) and a set of prohibitive elementary areas (PEA). Thinking like that, we can define different two-dimensional matrices to characterize each EA. Let consider the following matrices:

- Cover defining whether an EA requires coverage;
- Place defining whether in an EA we can place a node;
- CoverDepth defining the number of routers covering an EA

Therefore, an EA at position $(x, y)$ can be characterised by (1), (2) and (3):

$$Cover(x,y) = \begin{cases} 0 \rightarrow coverage\ not\ required \\ 1 \rightarrow coverage\ required \end{cases} \quad (1)$$

$$Place(x,y) = \begin{cases} 0 \rightarrow cannot\ place\ a\ node \\ 1 \rightarrow can\ place\ a\ node \end{cases} \quad (2)$$

$$CoverDepth(x,y) = \begin{cases} 0 \rightarrow no\ coverage \\ x \rightarrow covered\ by\ x\ routers \end{cases} \quad (3)$$

Fig. 1 illustrates the result of a decomposition of a region into a set of EA.

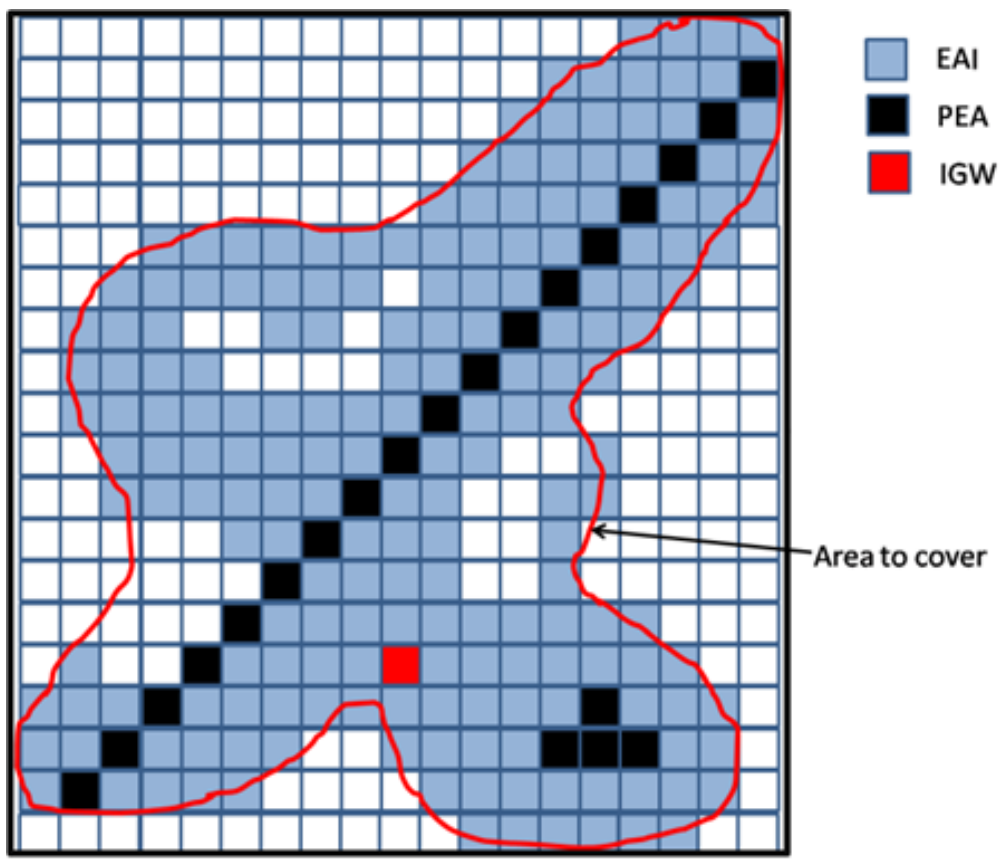

Fig. 1. An example of region decomposed in EA.

Since the population is not so dense like in urban regions, we consider a uniform repartition of clients in EAIs. That means each EAI has the same number of client. We consider routers to have the same radius ($r$). This radius is expressed in number of EA. For example $r$= 6 means that the radius stretches over 6 EAs.

Let p an EA at position $(x, y)$. If a MR is located in p, then the set of EA covered by this MR is given by (4).

$$\forall(a,b), (x-a)^2 + (y-b)^2 < r^2 \quad (4)$$

## 3.2. Problem Statement

One of the biggest concerns when deploying WMN in a rural region is the overall cost. This cost is influenced by the number of mesh router nodes. The more the region to cover is big, the more we need routers and the cost increases. So to minimize the cost, we need to cover only areas of interest. Therefore, the MR placement problem in rural regions can be described as the determination of minimum set of positions which maximizes the coverage of areas of interest, minimizes the coverage of optional areas while minimizing the number of MRs and ensuring the connectivity.

# 4. Placement Approach

## 4.1. Basic Metropolis Algorithm

Metropolis algorithm is a meta-heuristic designed to solve global optimization problems by finding a good approximation to the global optimum. Metropolis algorithm is a specialization of simulated annealing algorithm with a non-variant temperature T, used in the acceptability criteria. A pseudo code for metropolis algorithm is the following:

Algorithm1: Metropolis

Set T

```
S:=Initial Solution()
V:=Evaluate(S)
while(stopping condition not met) do
    St:=Generate(S)
    Vt:=Evaluate(St)
    if Accept(V, St, T ) then
        S:=St
        V:=Vt
    end if
endwhile
return S
```

## 4.2. Algorithm Particularization

The basic algorithm is particularized as follow: First we generate an initial solution and we evaluate it. Afterwards, we set the number of iteration of the algorithm called NumIter and also the maximum number of iteration allowed without amelioration of the solution called Stop. Therefore we select a router in each round, we simulate a movement and we check whether the movement is acceptable or not. If yes, we consider the movement and we reset stop since a movement has been accepted. We continue until NumIter=0 or Stop=0. Then we check whether the routers are connected and the coverage percentage of Area of Interest is greater or equal to a required percentage e. If yes, we save the present configuration, we remove one router and we restart the algorithm with this new number of routers. We continue until either the routers are no longer connected or the percentage is less than the required coverage e. Therefore, an optimal number and positions of mesh router nodes are given by the previous configuration. The workflow of this particularized algorithm is given in Fig. 2.

## 4.3. Algorithm Parameter

### 4.3.1. Initialization

The first step is to determine the initial number of routers for a given region. For that, we need first to determine the minimum number of router that allows full coverage of EAIs. This is calculated by dividing the total required area by the area covered by a router. Let $r$ be the radius of a router, the minimum number of router is given by (5).

$$nr_{min} = \lceil \sum Cover(x,y) \ /(r^2 \times 3.14) \rceil \quad (5)$$

Because this minimal number of routers may not ensure the coverage and the connectivity of the required areas (since routers should overlap to ensure the connectivity), we use a greater number so that the coverage and the connectivity can be ensured. Later this number is reduced while trying to keep the same coverage. However, a too great number at the beginning is not efficient. We choose an initial number of routers given by (6).

$$\mathbf{1.5} * \boldsymbol{nr_{min}} < \boldsymbol{nr_{init}} \leq \mathbf{2} * \boldsymbol{nr_{min}} \quad (6)$$

During this phase of initialization, routers are place randomly in the region, but only on areas of interest. For each router we randomly select an EA. We check if Cover(EA)=1 and Place(EA)=1 then the current router can be placed there. Otherwise, we continue by selecting another EA. The initialization ends when all routers are placed.

### 4.3.2. Movement

We define a set of movements and we move only one router at the same time. The movement is randomly selected. We define three hops for $r$; $r/2$; $r/4$ in different directions. We select randomly a distance and a direction and the movement from the current $EA_a$ to the new $EA_b$ is simulated if and only if Cover($EA_b$)=1 and Place($EA_b$)=1. This is always done to avoid placing router in optional area and consequently to reduce their coverage.

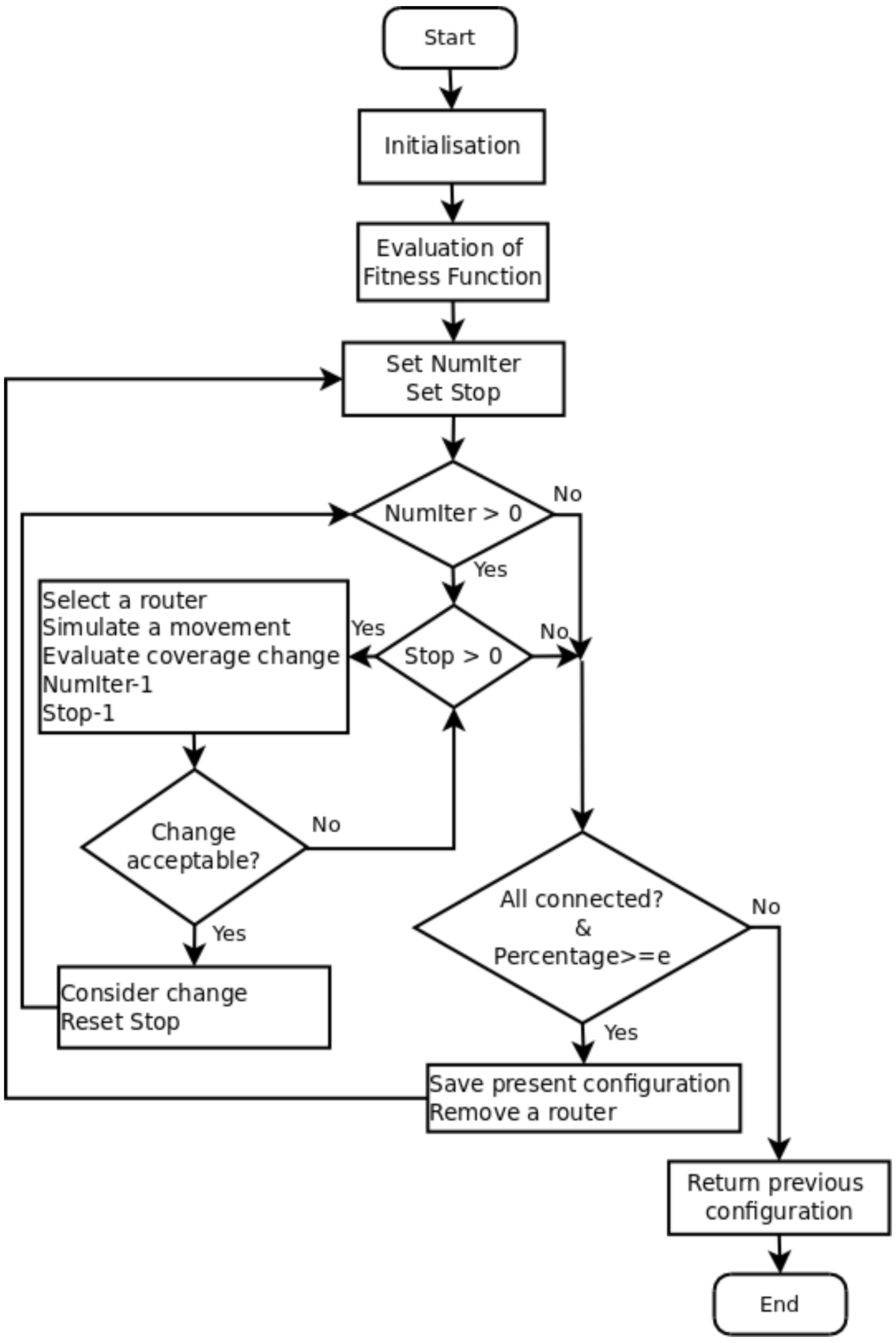

Fig. 2: Workflow of placement approach

### 4.3.3. Fitness function

To evaluate the fitness function, we count the number of EAI that are covered. This is done by (7) after the initialization.

$$f = \sum sign(CoverDepth.* \ Cover) \tag{7}$$

To be more efficient in the iteration phase, we calculate only the change in the coverage. Since we move only one router at the same time, we consider also only the EAs which are concerned by the movement. For example in Fig. 3, when moving a router from *a* to *b*, we decrease the coverage depth of router in position a and we increase the coverage depth of router in position b. Let $f_i$ be the value of the fitness function of the configuration i, $C_a$ the coverage of the selected router in previous position *a*, and $C_b$ the coverage of the selected router in new position *b*, the fitness function in configuration *i+1* is given by (8).

$$f_{i+1} = f_i + C_b - C_a \tag{8}$$

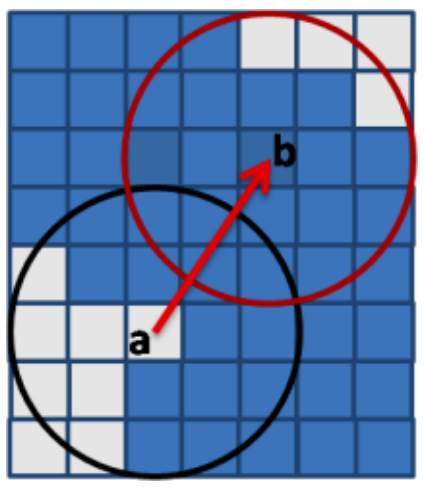

Fig. 3. Considering only EAs affected by a movement of a router.

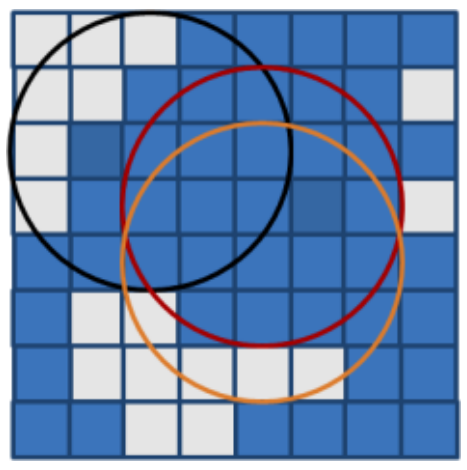

Fig. 4. Strategies for removing routers.

#### 4.3.4. Acceptability criteria

When $C_b \geq C_a$ , the coverage change is directly accepted. But when the coverage change is negative, to avoid local optimum, the change is accepted with a certain probability influenced by the temperature $T$. The acceptability criteria is given in (9) with $x$ a random number such that $0<x<1$.

$$rand(x) < exp(T * (C_b - C_a))$$

#### 4.3.5. Stopping condition

We stop looking for the improvement of a configuration when either the total number of iteration is reached (NumIter=0) or when the value of the fitness function does not improve after a certain number of iteration (Stop=0). Therefore we suppose having reached an optimal configuration.

#### 4.3.6. Nodes connectivity

At this step we check if each router is connected to at least one router while verifying that there exists no sub-network.

Let NotConnected be the set of not connected routers, Connected the set of connected routers, NumConnected the number of connected routers, nr the number of routers, fg a flag, DVec the difference between two points and Ctr the set of all router centres.

We consider initially that the first router belongs to Connected and the rest belong to NotConnected. We check that each router is connected to at least one router in Connected. When it is the case, the current router is removed from NotConnected and added to Connected and NumConnected is incremented. Two routers are connected if and only if the distance between their centres is less or equal to the sum of their radius.

The complete algorithm for checking the connectivity is the following:

```
  Algorithm 2: Checking connectivity
NotConnected := [0 ones(1,nr-1)]
Connected := [1 zeros(1,nr-1)]
NumConnected := 1
for ii:=1 to nr
    fg:=0
    for jj:=1 to nr
        if (NotConnected(jj)=1) then
            for kk:=1 to NumConnected
                DiffV:=Ctr(:,jj) - Ctr(:,Connected(kk))
                DistV:=DiffV'*DiffV
```

```
                    if (DistV<=4*r*r) then
                        fg:=1
                        NumConnected:=NumConnected+1
                   Connected(NumConnected):=jj
                   NotConnected(jj):=0
                   break
                    end
               end
          end
     end
     if (NumConnected==nr) then
          break
end
     if (fg==0) then
          break
      end
end
```

### 4.3.7. Optimal number of router

After ensuring a desired percentage of required coverage e, the next objective is to minimise the number of MR while keeping this percentage. We remove one router each time and perform movements with the rest. If the desired coverage percentage is satisfied, we continue to remove until we observe a variation of 1% or 2%. Therefore, we consider the previous number and placement of routers to be an optimal. To remove a router, three strategies can be used:

1) Remove router with minimum single-coverage: in Fig. 4 the orange router covers alone only one EAI (blue cells), the red router covers alone two EAIs and black router covers alone 5 EAIs. So, orange router should be removed;
2) Remove router with minimum coverage of EAI: in Fig. 4, the black router covers six optional EAs (white cells), the orange router covers four optional EAs and the red router does not cover optional EA. Therefore, the black router should be removed.
3) Remove circle with maximum over-coverage: always in Fig. 4, the black router has 10 EAIs over-covered, the orange router has 16 EAIs over-covered and the red one has 21 EAIs over-covered. So the red router should be removed.

Among these three strategies, the first appears to be the best.

## 5. Experiments and Results

### 5.1. Parameters Set up and Instances

To evaluate our proposed algorithm, we consider a grid of 200×200 with the radius of a router $r$=12. The unit is the size of an EA. If size (EA)=10m, the grid will be 2Km×2km=4km$^2$ and the radius $r$=120m. This is realistic since 802.11a routers have an outdoor theoretical transmission range of 120m, 802.11b routers a theoretical transmission range of 150m and 802.11n routers a theoretical transmission range of 250m. Another parameter is the temperature $T$=0.1. This value means that a movement of a router, which implies a great negative change in fitness function, has a weak probability to be accepted. The number of iteration is NumIter=4000, the number of iteration without improvement of the fitness function is Stop=500 and the number of initial routers is $nr_{Init}$=1.5*$nr_{min}$.

We generate three regions with different areas of interest, optional areas and prohibited areas as in Fig. 5a, Fig. 5b, and Fig. 5c. In these figures, white regions represent area of interest that should be covered and black regions represent optional areas. Prohibited areas are not directly seen on Figures because some areas of interest and optional areas can also be prohibited areas.

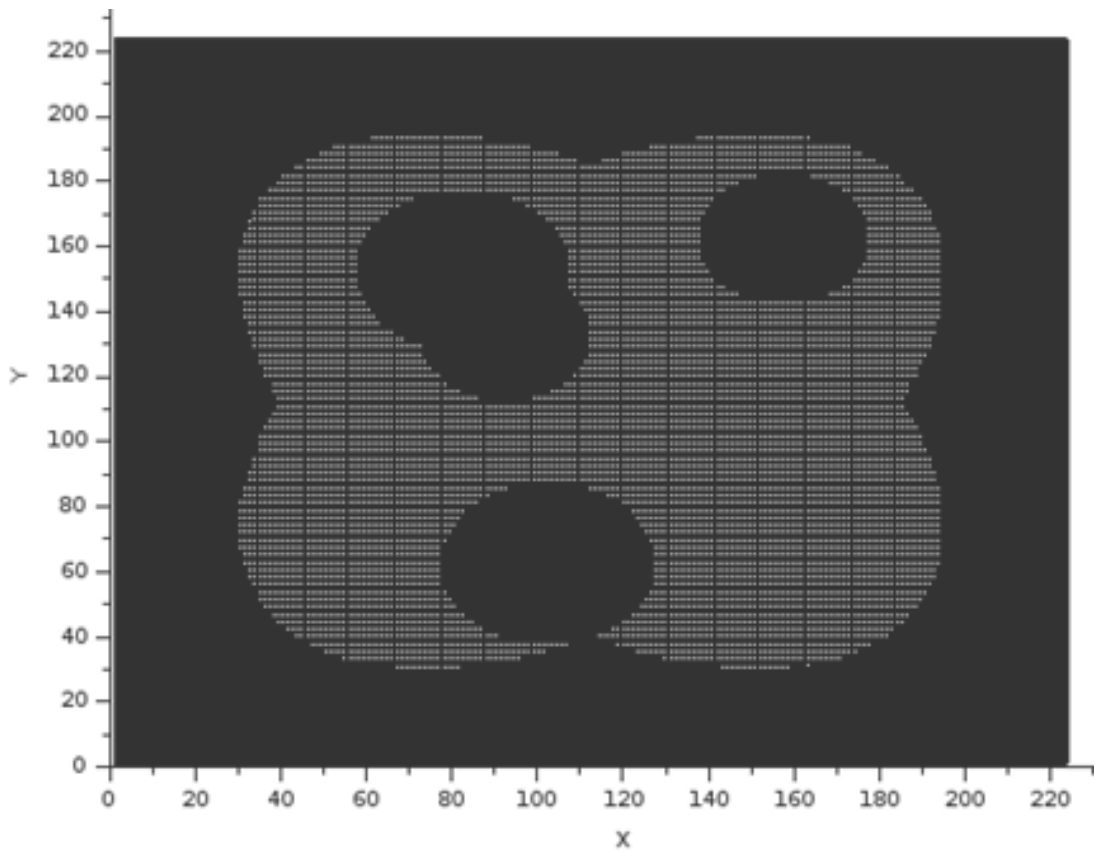

Fig. 5a. Areas to cover: instance 1

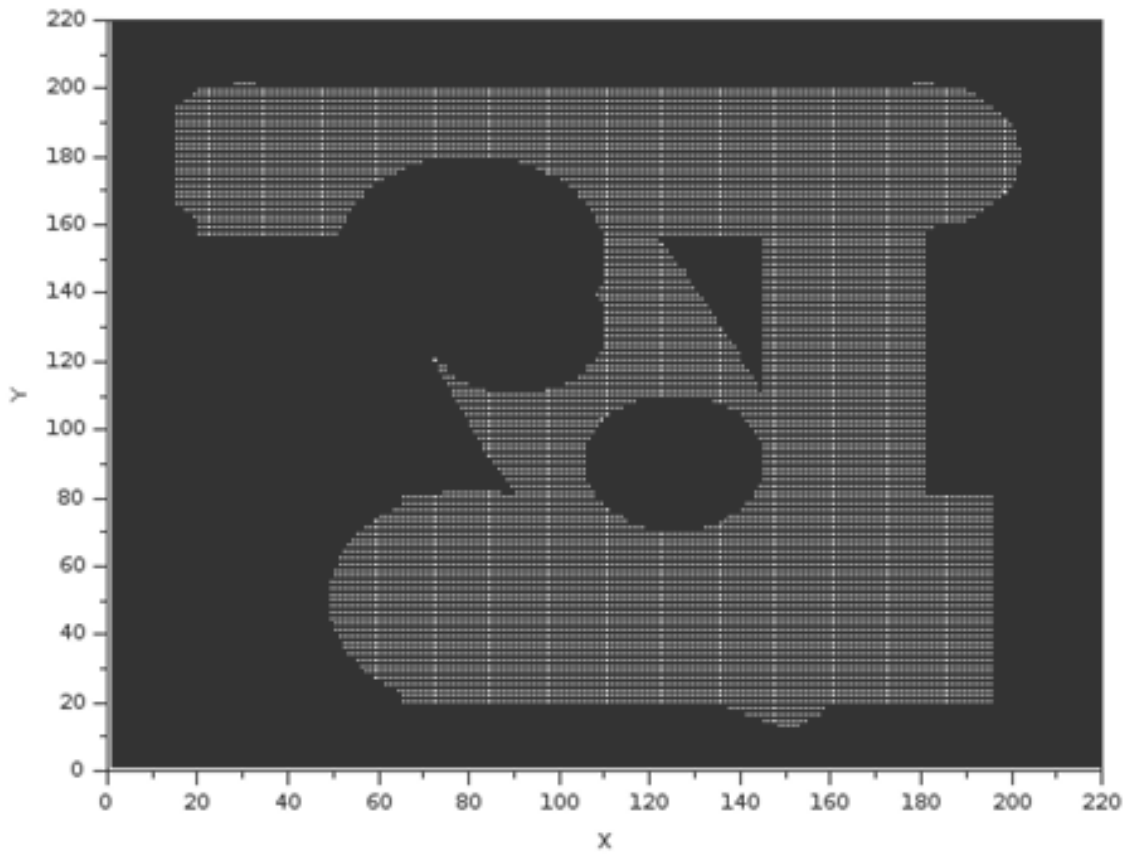

Fig. 5b. Areas to cover: instance 2

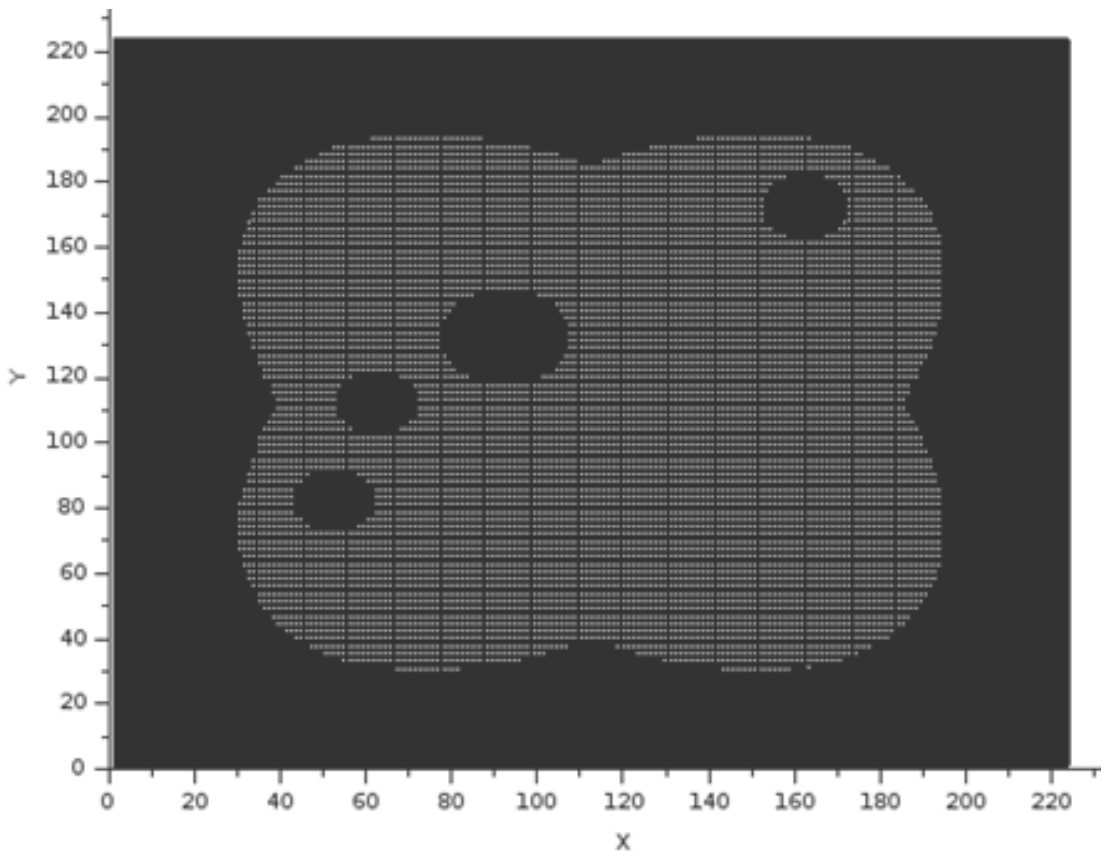

Fig. 5c. Areas to cover: instance 3.

## 5.2. Benchmark of Instances

For each instance, we run the algorithm three times. Table 1, Table 2, and Table 3 provide the results respectively for instance 1, 2, and 3. In these tables:

- $nr_{init}$ is the initial number of routers;
- $nr_{same}$ is the minimum number of routers that maintains the maximal coverage of area of interest from $nr_{init}$;
- $nr_{max-1}$ is the minimal number of routers for a decrease of 1% from the initial maximal coverage of area of interest;
- $nr_{max-2}$ is the minimal number of routers for a decrease of 2% from the initial maximal coverage of area of interest;
- $nr_{con}$ is the minimal number of routers which maintains the connectivity.
- $nr_{min}$ is the minimal number of routers calculated in (5).

For each run we determine $nr_{same}$ , $nr_{max-1}$ , $nr_{max-2}$ and $nr_{con}$. Determining $nr_{max-1}$ and $nr_{max-2}$ intends to reduce the number of routers while keeping a percentage close to the maximal obtained from $nr_{init}$. $nr_{same}$ aims to show that more than a certain number of routers, the coverage percentage of required areas provided by adding a new router is almost zero.

For these numbers of routers the coverage of areas of interest (Required Coverage in Table 1, 2, and 3) and the coverage of optional areas (Optional Coverage in Table 1, 2, and 3) is calculated, while checking whether all routers are connected.

Table 1. Results for instance 1

| Run | | Number of Router | All routers Connected? | Required Coverage | Optional Coverage |
|---|---|---|---|---|---|
| 1 | $nr_{init}$ | 65 | Yes | 97% | 22% |
| | $nr_{same}$ | 62 | Yes | 97% | 18% |
| | $nr_{max-1}$ | 59 | Yes | 96% | 16% |
| | $nr_{max-2}$ | 58 | Yes | 95% | 16% |
| | $nr_{con}$ | 53 | Yes | 91% | 13% |
| | $nr_{min}$ | 44 | No | 82% | 8% |
| 2 | $nr_{init}$ | 65 | Yes | 97% | 22% |
| | $nr_{same}$ | 62 | Yes | 97% | 18% |
| | $nr_{max-1}$ | 59 | Yes | 96% | 16% |
| | $nr_{max-2}$ | 58 | Yes | 95% | 15% |
| | $nr_{con}$ | 50 | Yes | 88% | 12% |
| | $nr_{min}$ | 44 | No | 82% | 8% |
| 3 | $nr_{init}$ | 65 | Yes | 97% | 19% |
| | $nr_{same}$ | 60 | Yes | 97% | 16% |
| | $nr_{max-1}$ | 58 | Yes | 96% | 15% |
| | $nr_{max-2}$ | 56 | Yes | 95% | 14% |
| | $nr_{con}$ | 48 | Yes | 88% | 12% |
| | $nr_{min}$ | 44 | No | 84% | 10% |

Table 2. Results for instance 2

| Run | | Number of Router | All routers Connected? | Required Coverage | Optional Coverage |
|---|---|---|---|---|---|
| 1 | $nr_{init}$ | 69 | Yes | 96% | 19% |
| | $nr_{same}$ | 66 | Yes | 96% | 19% |
| | $nr_{max-1}$ | 62 | Yes | 95% | 16% |
| | $nr_{max-2}$ | 60 | Yes | 94% | 14% |
| | $nr_{con}$ | 52 | Yes | 88% | 10% |
| | $nr_{min}$ | 46 | No | 82% | 7% |
| 2 | $nr_{init}$ | 69 | Yes | 97% | 21% |
| | $nr_{same}$ | 65 | Yes | 97% | 19% |
| | $nr_{max-1}$ | 63 | Yes | 96% | 16% |
| | $nr_{max-2}$ | 61 | Yes | 95% | 14% |
| | $nr_{con}$ | 48 | Yes | 85% | 9% |
| | $nr_{min}$ | 46 | No | 83% | 9% |
| 3 | $nr_{init}$ | 69 | Yes | 96% | 20% |
| | $nr_{same}$ | 65 | Yes | 96% | 17% |
| | $nr_{max-1}$ | 63 | Yes | 95% | 16% |
| | $nr_{max-2}$ | 60 | Yes | 94% | 16% |
| | $nr_{con}$ | 53 | Yes | 89% | 12% |
| | $nr_{min}$ | 46 | No | 83% | 9% |

Table 3. Results for instance 3

| Run | | Number of Router | All routers Connected? | Required Coverage | Optional Coverage |
|---|---|---|---|---|---|
| 1 | $nr_{init}$ | 80 | Yes | 97% | 16% |
| | $nr_{same}$ | 78 | Yes | 97% | 16% |
| | $nr_{max-1}$ | 74 | Yes | 96% | 15% |
| | $nr_{max-2}$ | 72 | Yes | 95% | 15% |
| | $nr_{con}$ | 53 | Yes | 82% | 8% |
| | $nr_{min}$ | 54 | Yes | 83% | 8% |
| 2 | $nr_{init}$ | 80 | Yes | 97% | 15% |
| | $nr_{same}$ | 75 | Yes | 97% | 13% |
| | $nr_{max-1}$ | 72 | Yes | 96% | 12% |
| | $nr_{max-2}$ | 69 | Yes | 95% | 12% |
| | $nr_{con}$ | 54 | Yes | 84% | 7% |
| | $nr_{min}$ | 54 | Yes | 84% | 7% |
| 3 | $nr_{init}$ | 80 | Yes | 97% | 17% |
| | $nr_{same}$ | 75 | Yes | 97% | 14% |
| | $nr_{max-1}$ | 71 | Yes | 96% | 12% |
| | $nr_{max-2}$ | 69 | Yes | 95% | 10% |
| | $nr_{con}$ | 52 | Yes | 83% | 5% |
| | $nr_{min}$ | 54 | No | 85% | 6% |

In Table 1 related to instance 1, all the three runs provide a percentage of 97% as required coverage with the initial number of router. But the third run provides the lowest optional coverage. Since the initial number of routers is the same for all the three runs, having the lowest percentage of optional coverage does not represent an advantage. However, when considering the number of routers equal to $nr_{max-2}$, the third run provides the more economical result with only 56 routers (1.27* $nr_{min}$), 95% as required coverage and only 14% of optional coverage. This last run also shows that the minimum number of routers that maintains the connectivity is 1.1* $nr_{min}$, with 88% as required coverage and 12% as optional coverage.

Related to instance 2, Table 2 shows, with the initial number of routers $nr_{init}$=69, a required coverage between 96 and 97%. The second run provides the best required coverage 97% with $nr_{init}$ routers but it has the greatest value of $nr_{max-2}$=61 when compared to 60 from run 1 and 3. In this case, adding a router just to increase the required coverage from 1% is not economical. In addition, the coverage of optional area is between 14% and 16%, which represents a significant reduction as it is the case in instance1. Therefore run 1 or 2, since they provide the lowest value for $nr_{max-2}$ (1.3* $nr_{min}$), are seen as the best solution.

In Table 3 from instance3, all the three runs provided a required coverage of 97% with the initial number of routers $nr_{init}$=80. The second run provided the more economical result with the lowest number of routers $nr_{max-2}$=69 (1.27* $nr_{min}$). As it is the case in instance1 and 2, we have small coverage percentage of optional area between 10% and 15%.

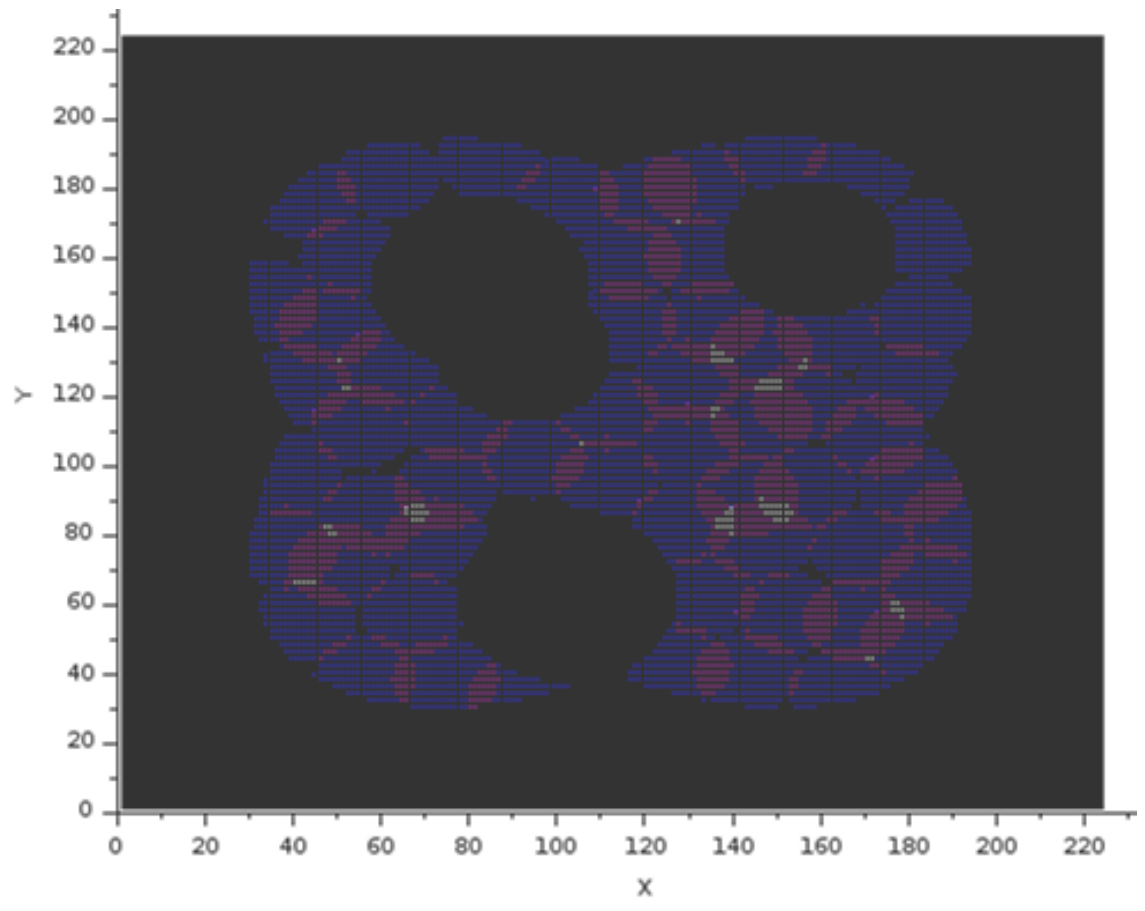

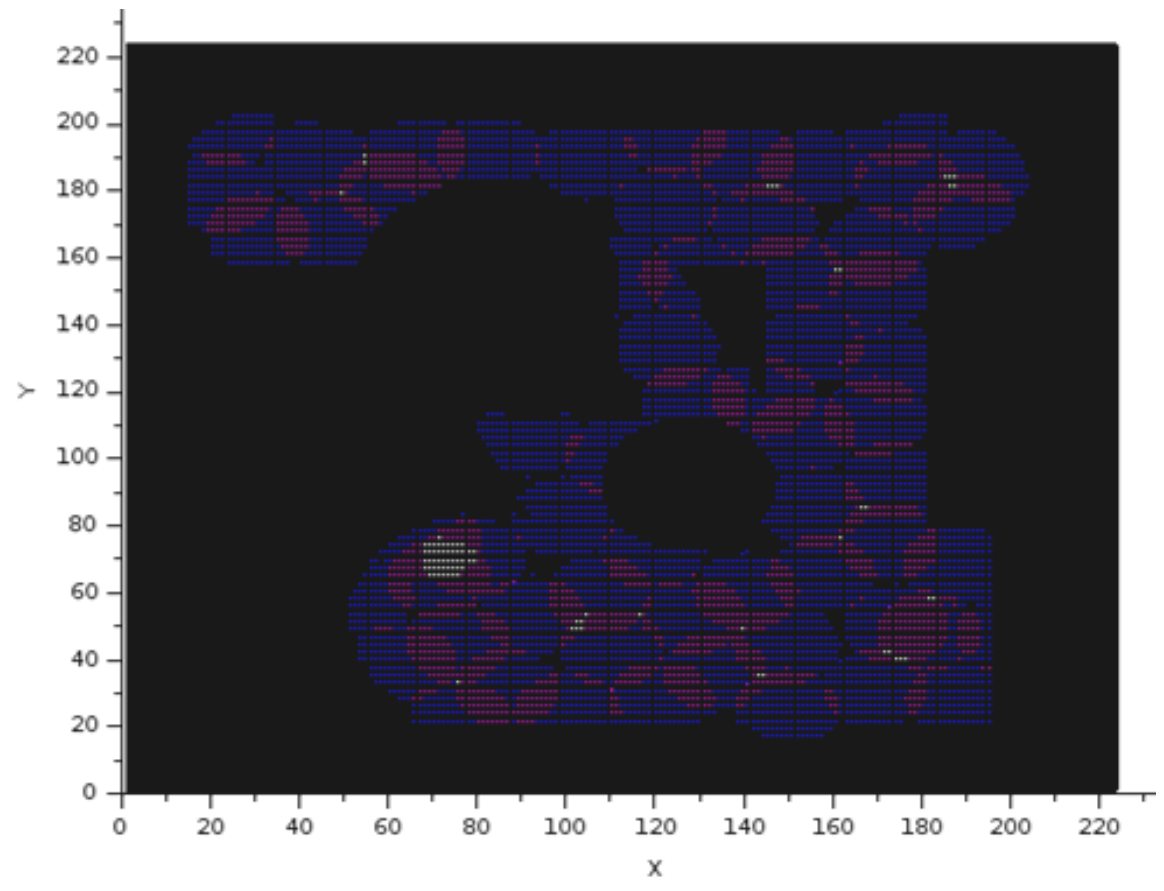

Fig. 6a. Covered Area of Instance1 with nrmax-2 routers Fig. 6b. Covered Area of Instance2 with nrmax-2 routers

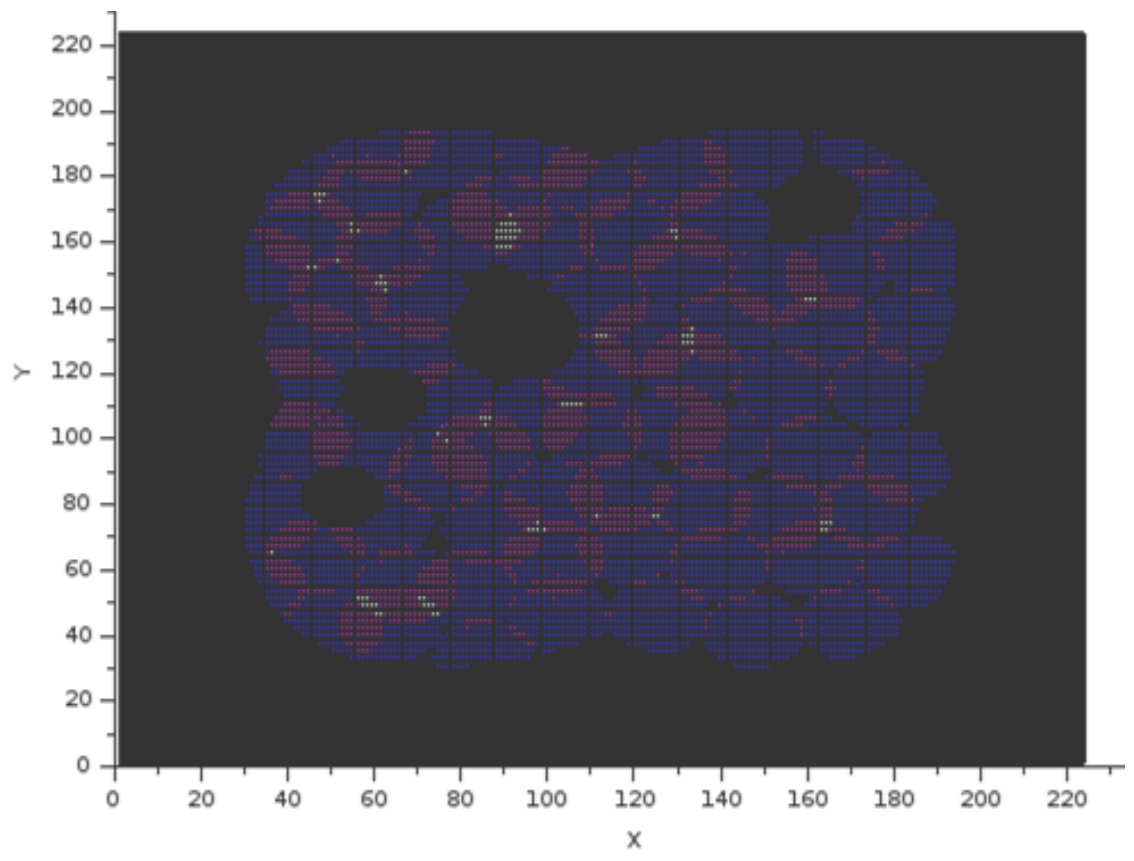

Fig. 6c. Covered Area of Instance3 with nrmax-2 routers.

Fig. 6a, Fig. 6b, and Fig. 6c show the placement of mesh router nodes in the areas to cover respectively for instances 1, 2, and 3. In these Figures, blue areas are covered by one router, red areas are covered by two

routers and white areas are covered by three routers. From these Figures, we observe that, despite the irregularity of areas to cover, this approach is efficient enough to avoid multi coverage while maintaining the connectivity.

### 5.3. Generalization of Results

From these three dissimilar instances and different runs of the algorithm, we can generalize some results. The first result is the efficiency of the approach to solve this problem of mesh router nodes placement in regions with different forms. In fact, in the three instances the average coverage is around 97% with $nr_{init}=1.5*nr_{min}$. Secondly, we observe a reduction of number of routers between 8% and 10% when decreasing the threshold coverage of required areas of 1% from the maximal percentage with $nr_{init}$ routers. Also, when considering an optimal number of routers $nr_{max-2}$ (with a decrease of 2% from the maximal coverage), we observe that this optimal number is around 1.3* $nr_{min}$, with a coverage percentage of optional areas less than 16%.

## 6. Conclusion

This paper has presented an approach based on Metropolis algorithm to solve the problem of mesh nodes placement in rural WMN. In this paper, we defined a network model and we clearly explained our approach based on the metropolis algorithm. We applied this particularised algorithm on three instances. Results from the experimentation showed the efficiency of this approach to solve the problem of mesh router nodes placement in rural areas while determining an optimal number of MRs. Indeed, the percentage of required coverage is around 97%. An optimal number of mesh router nodes with a decrease of 2% on required coverage is $nr_{max-2}=1.3*nr_{min}$, providing a coverage percentage of optional areas less than 16%.

In this work, we used a fixed temperature in the acceptability criteria as it is the case in Metropolis algorithm. It could be interesting to study whether the variation of the temperature improves the result. Also as future work, we should consider cases where an area of interest is disjointed from others; because this kind of situation usually results in a separated mesh network topologies. So, besides improving the algorithm in order to obtain a percentage very close to 100, with a number of routers less than $1.3*nr_{min}$, we will investigate on the case of disjointed areas of interest.

## Acknowledgment

This work is partially supported by a grant 2014-2015 from DAAD (Deutscher Akademischer Austauschdienst).

## References

[1] Akyildiz, I. F., Wang, X., & Wang, W. (2005). Wireless mesh networks: a survey. *Computer Networks, 47(4),* 445-487.

[2] Amaldi, E., Capone, A., Cesana, M., Filippini, I., & Malucelli, F. (2008). Optimization models and methods for planning wireless mesh networks. *Computer Networks, 52(11),* 2159-2171.

[3] Xhafa, F., Barolli, A., Sánchez, C., & Barolli, L. (2011). A simulated annealing algorithm for router nodes placement problem in wireless mesh networks. *Simulation Modelling Practice and Theory, 19(10),* 2276-2284.

[4] Fendji, J. L. E. K., Thron, C., & Nlong, J. M. (2014). Mesh router nodes placement in rural wireless mesh networks. In M. Sellami, E. Badouel, & M. Lo(Eds.), *Actes du CARI 2014 (Colloque Africain Sur LA Recherche en Informatique et Mathématiques Appliquées)* (pp. 265-272). Inria: Colloques CARI.

[5] Benyamina, D., Hafid, A., & Gendreau, M. (2012). Wireless mesh networks design — A survey. *Communications Surveys & Tutorials, 14(2),* 299-310.

[6] Benyamina, D., Hallam, N., & Hafid, A. (2008). On optimizing the planning of multi-hop wireless networks using a multi objective evolutionary approach. *Int. J. Commun*, *4(4),* 213-221.

[7] Robinson, J. (2009). Deployment and assessment of wireless mesh networks. PhD Thesis, Rice University, Houston Texas, USA.

[8] Wang, J., Xie, B., Cai, K., & Agrawal, D. P. (2007, October). Efficient mesh router placement in wireless mesh networks. *Proceedings of IEEE International Conference on Mobile Adhoc and Sensor Systems* (pp. 1-9). 2007.

[9] Xhafa, F., Sánchez, C., & Barolli, L. (2010, April). Genetic algorithms for efficient placement of router nodes in wireless mesh networks. *Proceedings of 24th IEEE International Conference on Advanced Information Networking and Applications (AINA)* (pp. 465-472). 2010.

[10] Xhafa, F., Sanchez, C., & Barolli, L. (2009, June). Adhoc and neighborhood search methods for placement of mesh routers in wireless mesh networks. *Proceedings of 29th IEEE International Conference on Distributed Computing Systems Workshops* (pp. 400-405). 2009.

[11] De Marco, G. (2009, September). MOGAMESH: A multi-objective algorithm for node placement in wireless mesh networks based on genetic algorithms. *Proceedings of 6th International Symposium on Wireless Communication Systems* (pp. 388-392). 2009.

[12] Franklin, A. A., & Murthy, C. S. R. (2007, November). Node placement algorithm for deployment of two-tier wireless mesh networks. *Proceedings of Global Telecommunications Conference* (pp. 4823-4827). 2007.

[13] Li, F., Wang, Y., Li, X. Y., Nusairat, A., & Wu, Y. (2008). Gateway placement for throughput optimization in wireless mesh networks. *Mobile Networks and Applications*, *13(1-2),* 198-211.

[14] Aoun, B., Boutaba, R., Iraqi, Y., & Kenward, G. (2006). Gateway placement optimization in wireless mesh networks with QoS constraints. *IEEE Journal on Selected Areas in Communications, 24(11),* 2127-2136.

[15] Muthaiah, S. N., & Rosenberg, C. (2008, June). Single gateway placement in wireless mesh networks. *Proceedings of 8th International IEEE Symposium on Computer Networks* (pp. 4754-4759).

[16] Zhou, P., Manoj, B. S., & Rao, R. (2007, October). A gateway placement algorithm in wireless mesh networks. *Proceedings of the 3rd International Conference on Wireless Internet* (p. 1).

[17] Younis, M., & Akkaya, K. (2008). Strategies and techniques for node placement in wireless sensor networks: A survey. *Ad Hoc Networks*, *6(4),* 621-655.

[18] Tan, H., Hao, X., Wang, Y., Lau, F., & Lv, Y. (2013). An approximate approach for area coverage in wireless sensor networks. *Procedia Computer Science*, *19*, 240-247.

**Jean Louis E. K. Fendji** was born in Douala, Cameroon in 1986. He received the B.Sc. and M.SC. degrees in computer science from University of Ngaoundéré, Cameroon, in 2007 and 2010, respectively.

He has been working as a scientist in the BMBF-Project CMR 10/P01 between the University of Ngaoundéré and the University of Bremen (2011-2013). He is currently an assistant lecturer at the University of Ngaoundéré and a Ph.D. candidate in computer science at the University of Bremen. Current research interests focus on optimisation techniques for the design of sustainable network and services.

**Chris Thron** was born in Boston, Massachusetts USA in 1957. He received a B.A. degree in mathematics from Princeton University in 1980, a Ph.D. degree in mathematics from the University of Wisconsin in 1985, and a Ph.D. degree in physics from the University of Kentucky in 1997.

He was a foreign expert (lecturer) at several universities in China from 1985-1990. He has also worked at a systems engineer for NEC, Motorola, and Freescale from 1995 to 2008, and has held academic positions at King College (Bristol, TN), Baylor University, and Austin Community College. Currently he is an assistant professor and the chair of the Mathematics and Sciences Department, Texas A&M –Central Texas (Killeen,TX). His Current research interests include algorithm design and optimization with applications to communications systems and epidemiology.

**Jean Michel Nlong** obtained his master degree in computer science from the University of Yaounde I, Cameroon, in 1998, and his Ph.D. degree from the same university in 2008, after a four-year stay at the Institut National Polytechnique Grenoble France, from 2002 to 2005, at the Laboratoire Infromatique et Distribution. His main interest is computer science, system and communications.

He has been working as an assistant lecturer at the University of Ngaoundere, Ngaoundere, Cameroon, since 1999, and became a senior lecturer in 2009. His actual research activities include peer-to-peer systems, mobile computing and security, Android OS, middleware.